\documentclass[12pt]{article}
\usepackage{graphicx,amsmath,bbm,amssymb,amsfonts,epsfig,here,psfrag}
\newcommand {\slsh} [1] {\not{\hbox{\kern-2pt${#1}$}}}

\def\drawbox#1#2{\hrule height#2pt
         \hbox{\vrule width#2pt height#1pt \kern#1pt
               \vrule width#2pt}
               \hrule height#2pt}

\def\Fund#1#2{\vcenter{\vbox{\drawbox{#1}{#2}}}}
\def\Asym#1#2{\vcenter{\vbox{\drawbox{#1}{#2}
               \kern-#2pt       
               \drawbox{#1}{#2}}}}

\def\fund{\Fund{6.4}{0.3}}
\def\asymm{\Asym{6.4}{0.3}}
\def\bfund{\overline{\fund}}
\def\basymm{\overline{\asymm}}

\newcommand {\beq} {\begin{equation}}
\newcommand {\eeq} {\end{equation}}
  \newcommand {\ber}{\begin{eqnarray*}}
  \newcommand {\eer} {\end{eqnarray*}}
\newcommand {\bea}{\begin{eqnarray}}
  \newcommand {\eea} {\end{eqnarray}}

\newcommand{\Dslash}{\,{\raise.15ex\hbox{/}\mkern-12mu D}}

\def\Acknowledgements{\bigskip  \bigskip {\begin{center} \begin{large}
              \bf ACKNOWLEDGMENTS \end{large}\end{center}}}

\begin{document}
\begin{titlepage}

\vskip 1cm
  
\centerline{{\Large \bf A Non-Supersymmetric Large-N 3D CFT}}
\vskip 0.4mm
\centerline{{\Large \bf And Its Gravity Dual}}

\vskip 1cm
\centerline{\large{  Adi Armoni {$^a$}  and Asad Naqvi $^{a,b}$}}

\vskip 0.5cm
\centerline{${}^a$ \it Department of Physics, Swansea University}
\centerline{\it Singleton Park, Swansea, SA2 8PP, UK}
\vskip 0.3cm
\centerline{${}^b$ \it Department of Physics, Lahore University of Management and Sciences,}
\centerline {\it Lahore, Pakistan}
\vskip 1cm

\begin{abstract}
We propose a three dimensional non-supersymmetric  theory that is conformal in the large N limit.
 In a certain well defined bosonic sub-sector of gauge invariant operators, this theory is planar equivalent to the theory recently proposed by Aharony, Bergman, Jafferis and Maldacena  as the theory on multiple M2 branes at an orbifold singularity.  We discuss the realization of the theory on a brane configuration of type 0B string theory. Moreover, we propose an 11D gravity dual, obtained by a  projection of M-theory on $AdS_4 \times S^7 / Z_k$. 

\end{abstract}

\end{titlepage}

\section{Introduction}
\label{introduction}
\noindent

Recently Bagger and Lambert (BL) constructed a Lagrangian of a  three dimensional conformal field theory with ${\cal N}=8$ supersymmetry, which was conjectured to live on the worldvolume of multiple M-theory membranes \cite{BL,Gustavsson}. The basic element of their construction is a 3-algebra which is endowed with a totally anti-symmetric trilinear product and a positive definite inner product, which was inspired by the work of  Basu and Harvey \cite{Basu:2004ed}. The only concrete realization of this algebra is a four dimensional algebra with an SO(4) action \cite{Papadopoulos:2008sk,Gauntlett:2008uf} (see also \cite{Friedmann:2008rh}). This original BL SO(4) theory was re-written as an ordinary $SU(2)\times SU(2)$ Chern-Simons theory  with matter \cite{VanRaamsdonk:2008ft}.

More recently Aharony, Bergman, Jafferis and Maldacena (ABJM)\cite{Aharony:2008ug} (see also \cite{Benna:2008zy},\cite{Ahn:2008ya}) generalized the analysis of BL by considering a $SU(N) \times SU(N) $ Chern-Simons theory with matter. In the case $N=2$, there is a 3-algebra structure that is absent for $N > 2$. This theory lives on the worldvolume of $N$ M2-branes on a $C^4/Z_k$ orbifold singularity. The parameter $k$ corresponds to the level of the Chern-Simons theory. ABJM also found the 11D supergravity (M-theory) dual to the  three dimensional conformal field theory. Following \cite{Aharony:2008ug}, a number of papers \cite{many} which discuss various aspects of this theory have appeared.
 
In this short note we propose a new non-supersymmetric field theory in three dimensions that is conformal to leading order in the large-$N$ limit. This field theory belongs to a class of `orientifold field theories' \cite{sag,Armoni:2003gp} (similar to `orbifold field theories' \cite{Kachru:1998ys}). It can be obtained by a certain `orientifold projection' in ABJM field theory, or alternatively its spectrum can be read from a type 0 brane configuration. Both  the field theory and the string theory analyses suggest that the theory becomes equivalent to the ABJM theory theory at large $N$\footnote{A necessary and sufficient condition for `orientifold planar equivalence' to hold is an unbroken charge conjugation symmetry \cite{Kovtun:2004bz,Unsal:2006pj,Armoni:2004ub} (this condition is equivalent to the absence of closed string tachyons in the gravity dual \cite{Armoni:2007jt}).}. In particular, the field theory is expected to be conformal in the planar limit \cite{Gaiotto:2007qi}.

This paper is devoted to the investigation of this field theory. In section \eqref{CFT} we introduce the field theory and provide field theory arguments in favour of its planar equivalence with the ABJM theory. In section \eqref{orientifold} we review the orientifold version of ${\cal N}=4$ super Yang-Mills, and discuss its string theory realization via the Sagnotti model \cite{sag} and its gravity dual. In section \eqref{dual} we discuss the string theory realization and the M-theory bulk dual of the proposed  three dimensional CFT. Finally section \eqref{conclusions} is devoted to a short discussion.    

\section{A Non-Supersymmetric  three dimensional CFT}
\label{CFT}
\noindent

In this section we  propose a  three dimensional non-supersymmetric CFT. We argue that
in the large-$N$ limit the theory becomes planar equivalent to the ABJM theory in a certain well defined bosonic sector of gauge invariant operators.

The theory that we consider belongs to a class called `orientifold field theories'. The matter content of the theory can be read from a brane configuration of type 0A string theory, or it can be obtained by a `projection' in field theory. The result is a theory where the bosons transform as in the `parent' supersymmetric theory, but the fermions transform differently. The Lagrangian of the two theory is identical, namely it contains the same interaction terms.

The ABJM model is a level $(k,-k)$ $U(N)\times U(N)$ Chern-Simons theory, with four bi-fundamental bosons ($Z,Z^\dagger$) and their fermionic superpartners ($\zeta,\zeta ^\dagger$) (we use the notation of Benna et.al. \cite{Benna:2008zy}). In addition, there are auxiliary fields, transforming in the adjoint representation of each gauge group. The fermionic auxiliary fields are denoted by $\chi$ and $\hat \chi$ while the bosons are denoted by $D, \sigma, \hat D, \hat \sigma$.

The matter content of the ABJM theory is summarized in table \eqref{table-ABJM} below.

\begin{table}[H]
\begin{displaymath}
\begin{array}{l c@{ } c@{ } c}
& \multicolumn{1}{c@{\times}}{U(N)}
& \multicolumn{1}{c@{}}{U(N)} \\
\hline
A_\mu  & adj.  & 1 \\
D & adj. & 1 \\
\sigma & adj. & 1 \\
\chi & adj. & 1 \\
\hat A_\mu  & 1 & adj. \\
\hat D & 1 & adj. \\
\hat \sigma & 1 & adj.  \\
\hat \chi & 1 & adj. \\
\hline
 \rm Z & \bfund & \fund  \\
 \rm Z^\dagger & \fund &  \bfund  \\
 \zeta & \bfund & \fund  \\
 \zeta ^\dagger & \fund &  \bfund \\
\end{array}
\end{displaymath}
\caption{The field content of the ABJM theory.}
\label{table-ABJM}
\end{table}

We now describe the matter content of the non-supersymmetric `orientifold theory'. It is obtained by changing the representations of the fermions of the ABJM theory. We replace the adjoint `gluinos' $\chi$ by antisymmetric (or equivalently, symmetric) Dirac fermions. The interaction term $Z\bar \chi \zeta$ forces us to change the representation of the bi-fundamental fermions: in the new theory they transform either in the $(\fund, \fund)$ or in the $(\bfund , \bfund)$ of $U(N)\times U(N)$. The matter content of the non-supersymmetric `orientifold' theory is summarized in table \ref{table-orientifold} below.

\begin{table}[H]
\begin{displaymath}
\begin{array}{l c@{ } c@{ } c}
& \multicolumn{1}{c@{\times}}{U(N)}
& \multicolumn{1}{c@{}}{U(N)} \\
\hline
A_\mu  & adj.  & 1 \\
D & adj. & 1 \\
\sigma & adj. & 1 \\
\chi & \asymm + \basymm & 1 \\
\hat A_\mu  & 1 & adj. \\
\hat D & 1 & adj. \\
\hat \sigma & 1 & adj.  \\
\hat \chi & 1 & \asymm + \basymm \\
\hline
 \rm Z & \bfund & \fund  \\
 \rm Z^\dagger & \fund &  \bfund  \\
 \zeta & \fund & \fund  \\
 \zeta ^\dagger & \bfund &  \bfund \\
\end{array}
\end{displaymath}
\caption{The field content of the `orientifold' theory.}
\label{table-orientifold}
\end{table}

The resulting Lagrangian after integrating out the auxiliary fields is almost identical to the original ABJM theory, except that the representation of the fermions $\zeta, \zeta ^\dagger$ is different. Due to the difference in the representations of the fermions, the set of gauge invariant operators is purely bosonic. The reason is as follows: under a gauge transformation $U, \hat U$, the fermion $\zeta$ transforms as $\zeta \rightarrow U \zeta \hat U$ (and similarly $\zeta ^\dagger  \rightarrow \hat U ^\dagger  \zeta ^\dagger U ^\dagger$), and therefore a gauge invariant operator includes an even number of fermions. 

We now show that the `orientifold' theory is planar equivalent to the ABJM theory. In order to obtain an intuition about the non-perturbative equivalence, let us begin with a simple perturbative example. Of course, the theories admit a ${1 \over N}$ expansion in the `t Hooft limit with $\lambda \equiv N/k$ fixed. At each order in $N$, there is a perturbative expansion only when the Chern-Simons level is large enough $k \gg N$ and the `t Hooft coupling $\lambda$ is small \cite{Aharony:2008ug}. A typical (vacuum) planar diagram of a theory with a $U(N)\times U(N)$ gauge group is presented in figure \eqref{feynman} below. In the `t Hooft notation gauge fields are represented by blue-blue and red-red lines and fermions by blue-red (red-blue) lines. The operation $(\bfund, \fund) \rightarrow (\fund , \fund)$,  $(\fund, \bfund) \rightarrow (\bfund , \bfund)$, corresponds in the `t Hooft notation to reversing the arrow on one of the double lines that represent the fermion. When it is performed on the graph (b), it leads to the graph (c).  Evidently, this operation does not change the value of the Feynman graph. In general any vacuum planar graph of the ABJM theory can be drawn on a sphere, with blue and red regions, separated by fermionic loops. Reversing the arrows on, say, all the blue lines, leads to the non-supersymmetric `orientifold theory'. Such an operation does not change the value of the graph. Thus, at large-$N$ the susy and non-susy theories are equivalent to all orders in perturbation theory in $\lambda$ (when such an expansion exists).

\begin{figure}[ht]
\centerline{\includegraphics[width=3in]{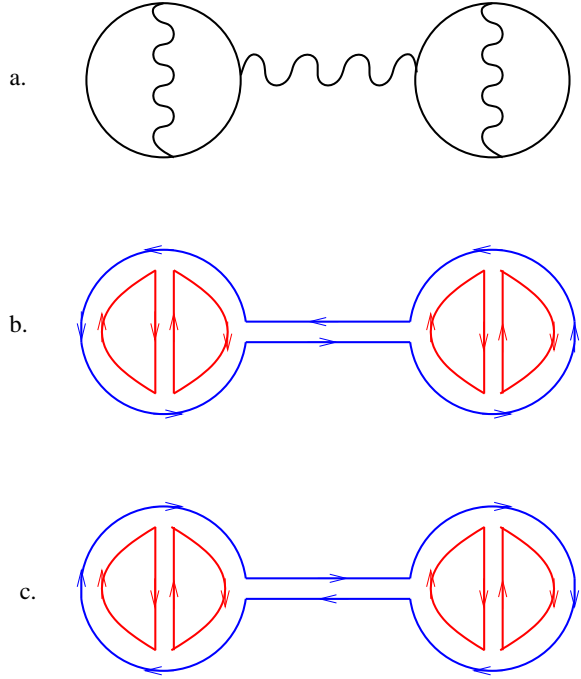}}
\caption{(a) A typical vacuum diagram. (b) A planar contribution in the supersymmetric theory. (c) The same planar graph in the non-supersymmetric theory. The graph (c) is obtained from the graphs (b) by reversing the arrows on the blue lines.} \label{feynman}
\end{figure}

We can also show the equivalence at the non-perturbative level by the following argument. This is a straightforward generalization of the proof of \cite{Armoni:2004ub}. Since fermions couple quadratically to other fields in the Lagrangian, ${\cal L}_{\rm int.} = \bar \zeta {\cal O} \zeta$, it is possible to integrate over the fermions. The form of the partition function, after the integration is
\beq
{\cal Z} = \int DA_\mu D\hat A_\mu D Z D Z^\dagger \exp -S_{\rm bosonic} \det {\cal O} \, .
\eeq
By using the worldline formalism \cite{Strassler:1992zr} it is possible to express
the fermionic determinant in terms of Wilson loops,
 \beq
 \det {\cal O} = \exp \Gamma \, ,
\eeq
 where the bosonic part of $\Gamma$ is given by
\beq
\Gamma = \int Dx ^\mu \exp (-S[x ^\mu, Z, Z^\dagger]) W \hat W \, , \label{gamma}
\eeq
and
\beq
W = {\rm tr} \exp i\int A ^a  _\mu t^a dx^\mu \, , \,\,\,\,\,\,\, \hat W={\rm tr} \exp i\int \hat A ^a  _\mu t^a dx^\mu  \, .
\eeq   
The difference between the non-supersymmetric theory and the ABJM theory is that in the ABJM theory the fermions transform in the bi-fundamental representation. Thus the partition function of the ABJM theory takes the same form, except that in eq.\eqref{gamma} $W$ has to be replaced by $W ^\dagger$. In general such a replacement changes the theory and this is why the `orientifold' theory is not equivalent to the ABJM theory at finite-$N$. At large-$N$, however, gauge invariant operators factorize. For example,
\beq
\langle W \hat W \rangle \rightarrow 
\langle W  \rangle 
\langle \hat W \rangle \, .
\eeq 
Moreover,  $\langle W  \rangle =\langle W^\dagger  \rangle $  if charge conjugation is not spontaneously broken . Therefore at large-$N$ 
\beq
\langle W \hat W \rangle = \langle W^\dagger \hat W \rangle \, . 
\eeq 
The above example demonstrates the large-$N$ relation between the non-supersymmetric theory and the ABJM theory. To complete the argument, note that the fermionic determinant can be written as a sum of products of Wilson loops
\beq
\det {\cal O} = \sum \prod W\hat W \, , 
\eeq
(or $\sum \prod W ^\dagger \hat W$ in the susy theory). It is possible to show \cite{Armoni:2004ub} that in the planar limit all connected Green functions made by Wilson-loops coincide, namely
\beq
\langle \prod W \hat W \rangle _{\rm conn.} = \langle \prod W^\dagger \hat W \rangle _{\rm conn.} \, . 
\eeq

 A different proof of `orientifold planar equivalence' was presented in \cite{Kovtun:2004bz,Unsal:2006pj}. Start with an $SO(2N)\times SO(2N)$ ABJM theory. Consider the matrix $J=i \sigma _2$ acting on $N\times N$ blocks of an $SO(2N)$ group. A supersymmetric projection, that leads to the $U(N)$ ABJM theory is obtained as follows: $A_\mu = J A_\mu J^t$, $\hat A_\mu = \hat J \hat A_\mu \hat J^t$,$Z = J Z \hat J^t$, $\zeta = J \zeta \hat J^t$. The non-supersymmetric orientifold theory is obtained by the projection: $A_\mu = J A_\mu J^t$, $\hat A_\mu = \hat J \hat A_\mu \hat J^t$,$Z = J Z \hat J^t$, $\zeta = -J \zeta \hat J^t$. The two resulting theories are planar equivalent to the `parent' $SO(2N)\times SO(2N)$ supersymmetric theory in the common `untwisted' sector: the sector of gauge invariant operators which are neutral under the two projections.

Both approaches, \cite{Armoni:2004ub} and  \cite{Kovtun:2004bz,Unsal:2006pj}, lead to the same conclusion: planar equivalence holds if and only if charge conjugation is not spontaneously broken.  

To conclude, we presented a non-supersymmetric theory with a partition function that becomes equivalent to the ABJM partition function at large-$N$. Therefore the theory is expected to be conformal at large-$N$. Note that the equivalence between the theories holds only in the sector of bosonic gauge invariant operators. The orientifold theory does not contain fermionic gauge invariant operators.

A large-$N$ CFT may admit a gravity dual. In section \eqref{dual} we will present a candidate for the dual of the `orientifold' theory.

\section{Orientifold Theories and Their Relation to String Theory}
\label{orientifold}
\noindent

In order to obtain a better understanding of orientifold field theories let
us discuss their string theory origin.

 Consider critical type 0B string theory in 10 dimensions. In \cite{sag},  Sagnotti introduced a special orientifold projection $\Omega^\prime= \Omega (-)^{f_R}$, where $f_R$ is the right moving fermion number on the world sheet.  This operation leads to a tachyon free non-supersymmetric string theory (called type $0^\prime$B) in which the massless modes in the closed string sector consist of a dilaton, a graviton, and a R-R zero-form, two-form and a self-dual four form. Note that the NS-NS two form is projected out. In order to cancel the R-R tadpole one has to introduce $32$ D9 branes that give rise to a $U(32)$ gauge theory with a two-index anti-symmetric Dirac fermion. It is instructive to think about the gauge theory as a hybrid of bosons from a $U(N)$ theory and fermions from a $SO(N)$ theory. Indeed, only fermions appear in the open string channel of the M\"obius amplitude. Type $0^\prime$B string theory is reviewed in \cite{DiVecchia:2005vm}.

The simplest four dimensional `orientifold field theory' is obtained by placing an $O^\prime3$ plane on top of a stack of $N$ D3 branes in type 0B string theory. The gauge theory on the branes is a $U(N)$ gauge theory with six real adjoint scalars and four Dirac fermions in the anti-symmetric representation. This theory was first introduced and discussed in \cite{Angelantonj:1999qg}. The theory becomes conformal at large-$N$ and in this limit it becomes equivalent to ${\cal N}=4$ $U(N)$ super Yang-Mills theory, in the common sector of bosonic gauge invariant operators.
Let us discuss the gravity dual of this theory. The large-$N$ equivalence with  ${\cal N}=4$ SYM suggests a duality with 10D type IIB supergravity on $AdS_5 \times S^5$.

Despite of the presence of the orientifold plane, the dual background cannot be $AdS_5 \times RP^5$ \cite{Witten:1998xy}, since the moduli space of the theory spanned by the six adjoint scalars of the $U(N)$ theory corresponds to $S^5$ and not $RP^5$. The dual does not consist of all the supergravity modes: since the gauge theory does not contain fermionic gauge invariant operators, the gravity dual cannot contain fermions either. This is in agreement with the absence of the NS-R sector in the Sagnotti model. Furthermore, the NS NS 2 form should be projected out from the bulk theory. To see that this is the case, notice that in the usual correspondence between type IIB on $AdS_5 \times S^5$ and ${\cal N}=4$ SYM, the NS NS B-field (after Kaluza-Klein) reduction on the $S^5$ couples to bosonic operators that contain and unequal number of left-handed and right-handed Weyl fermion, {\em i.e.} operators with a non-trivial `fermion number' charge. 
For example, the Kaluza-Klein tower that is obtained by Kaluza-Klein reducing the $B_{ab}$ (where $a$ and $b$ are indices on the $S^5$) couples to operators of the form
\beq
{\cal O} \sim \epsilon^{\alpha \beta} {\rm Tr} (\lambda^A_\alpha \lambda^B_\beta \phi^k)
\eeq
where $\alpha$ and $\beta$ are Weyl spinor indices and $A$ and $B$ label the fundamental representation of the SU(4) R-symmetry. Here, $\lambda$ is the adjoint fermion and $\phi$ is the adjoint scalar. These operators are not gauge invariant after we make  the `orientifold' projection and the left handed spinor transforms as an anti-symmetric product of two fundamentals and the right handed spinor transforms as an anti-symmetric 
product of two anti-fundamentals. Thus operators with an unequal number of left and right handed Weyl spinors will be projected out, which exactly couple to the NS-NS B-field in the bulk. 
 Therefore, our conjecture for the bulk dual theory to the planar  limit of the theory obtained from an `orientifold' projection of ${\cal N}=4$ theory is a projection of type IIB on $AdS_5 \times S^5$ that removes the fermions and the NS-NS B-field. There is no known closed string theory that has a massless spectrum that is the bosonic truncation of type IIB without the NS NS B-field (note that type $0^\prime$B has an open-string sector giving rise to a $U(32)$ gauge symmetry). 
 However, we emphasize that the dual is purely classical, since it provides the bulk description of the planar theory.  At non-planar level, we expect conformal invariance to be broken. This is clear from our field theory analysis, because the equivalence only holds at the planar level. Correspondingly, in the bulk, quantum effects will deform the AdS background, presumably to a space with string scale curvature. Therefore, while there is no known string theory with the massless spectrum of our proposed dual, the field theory at finite $N$ defines a string theory at the quantum level.

\section{The Gravity Dual of the Non-Supersymmetric three dimensional CFT}
\label{dual}
\noindent

The brane configuration in type 0B string theory that gives rise to the CFT of section \eqref{CFT}
 consists of $N$-D3 branes along the 012 directions and a compact direction 6, an NS5 brane along 012345 and a tilted $(1,k)$ 5-brane, as in \cite{Aharony:2008ug}. In addition we place an $O^\prime 3$ plane on top of the D3 branes. Note that the bulk tachyon does not couple to the D3 branes, due to presence of the $O^\prime 3$ plane. The configuration is depicted in figure \eqref{configuration} below.

\begin{figure}[ht]
\psfrag{NS5}{{\bf NS5}}
\psfrag{N D3 Branes +}{{\bf N D3 branes +}}
\psfrag{O'3 Plane}{{$\mathbf {O^\prime 3 ~plane}$}}
\psfrag{(1,k) 5}{$\mathbf{(1,k)~5 ~brane}$}
\centerline{\includegraphics[width=4in]{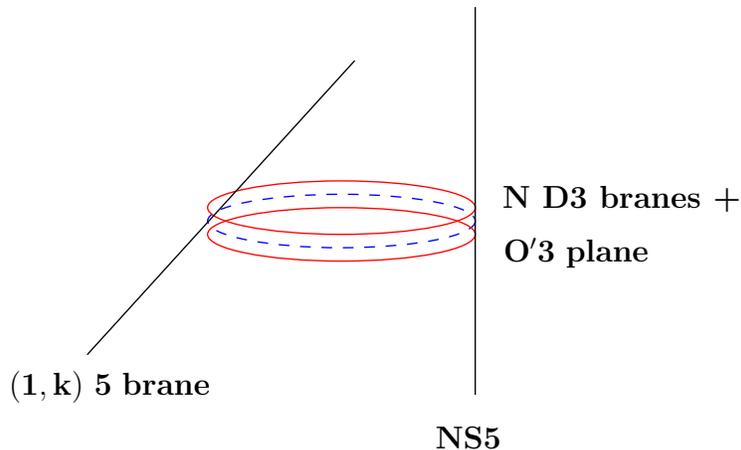}}
\caption{The type 0B brane configuration that gives rise to the `orientifold' CFT. The red (solid) lines are D3-branes. The blue (dashed) line is the $O^\prime 3$ plane.} \label{configuration}
\end{figure}

In particular, in the presence of the $O^\prime 3$ plane, bosons transform as in the supersymmetric theory, whereas the representations of the fermions become $(\fund,\fund)$ and $(\bfund,\bfund)$. 

A very similar brane configuration gives rise to an `orientifold' version of SQCD \cite{Armoni:2008gg}. In that case it is possible to `integrate out' the 5-branes and to obtain a description of the field theory in non-critical type $0^\prime$ string theory. In this description the bulk does not contain a tachyon. 
It would be interesting to obtain a description of either the original ABJM theory or its orientifold version via non-critical strings.   

Since the bulk tachyon does not couple to the field theory, we expect a dual which does not contain a tachyon.  This is essential for planar equivalence to hold: a tachyonic mode would invalidate the equivalence \cite{Armoni:2007jt}. We propose that the gravity dual is the bosonic truncation of M-theory on $AdS_4 \times S^7/Z_k$. This is due to the absence of fermionic gauge invariant operators in the field theory side. If we consider the $S^7$ as a U(1) fibration over $CP^3$, the $Z_k$ orbifold acts only on the $U(1)$ fiber. In the large $k$ limit,  in the supersymmetric setup, we obtain a type IIA theory by reducing M-theory on the $U(1)$ fiber. To go from type IIA to type 0A, we project out the fermions and the NS NS B-field. Therefore, in the dual description of the non-supersymmetric `orientifold' theory, we need to project out the component of the three form with one leg in the U(1) fiber direction, {\em i.e.} if we label the $U(1)$ fiber with the coordinate $\theta$, then we project out $A_{MN\theta}$, where $M$ and $N$ are 11 dimensional indices. This bulk theory, at the classical level, is dual to the planar limit of the `orientifold' theory. 

We  conclude this section with a comment about planar equivalence and the gravity dual. It is necessary that the gravity solution of the `orientifold' theory is identical to bosonic part the gravity solution corresponding to the parent theory \cite{Armoni:2007jt}. The absence of the bulk fermions, however, generically allows for more solutions. In particular when the gauge theory is formulated on a space with a non-trivial $S^1$, with {\em anti-periodic} boundary conditions for the gauge theory fermions, the gravity dual of the `orientifold' theory admits an AdS black-hole solution in which the boundary $S^1$ shrinks in the bulk (in the Euclidean framework). Fermions in the bulk can only have anti-periodic boundary conditions along a shrinking circle. If {\em periodic} boundary conditions  are imposed for the gauge theory fermions and the compactification radius is small the 'orientifold' theory admits a description in terms of an AdS black-hole, whereas the supersymmetric theory admit a description in term of thermal AdS and therefore planar equivalence does not hold. On $R^4$, however, the AdS solution is unique.

\section{Conclusions}
\label{conclusions}
\noindent

In this short note we presented a new  three dimensional field theory. We argued that in the large $N$ limit, this becomes equivalent to a sector of the ABJM theory.

The theory is conjectured to be dual to a certain truncation of M-theory on
$AdS_4 \times  S^7/Z_k$. 
The AdS/CFT duality tells us that this CFT lives on a collection of $N$ membranes, placed on a $C^4/Z_k$ orbifold singularity. If so, there should be a way to lift type $0^\prime$ string theory to M-theory (more precisely, to lift the corresponding 10D gravity theory to 11 dimensions). While at present it is not known how to perform this lift\footnote{A lift of type 0A to M-theory was conjectured in \cite{Bergman:1999km}.}, we hope that our results will shed light on this problem. In particular, in the supergravity limit of M-theory, the lifted theory should be obtained by the truncation discussed in the previous section.

\Acknowledgements

We thank O. Aharony, C. Hoyos and E. Imeroni for useful discussions. We are supported by the STFC (PPARC) advanced fellowship award.

\end{document}